\documentclass[pre,aps,twocolumn,amsfonts,showpacs]{revtex4}
\usepackage{amscd}
\usepackage{amsmath}
\usepackage{amssymb}
\usepackage{amsthm}
\theoremstyle{break}

\theoremstyle{change}
\begin{document}
\title{Some formal properties on superstatistics and superposition of statistical factors}
\author{{\sc Takuya Yamano}
\footnote{Division of Social Sciences, International Christian University, Osawa, Mitaka, 
Tokyo 181-8585 Japan (from April 2005).}}
\email{yamano@cosmos.phys.ocha.ac.jp}
\affiliation{Department of Physics, Ochanomizu University, 2-1-1 Otsuka, Bunkyo-ku, 
Tokyo 112-8610, Japan}
\date{\today}

\begin{abstract}
By focusing on the interchangeable role in a generating function (i.e., $\beta \leftrightarrow E$ 
in the Laplace transform), the superstatistics proposed by Beck and Cohen can be viewed as a 
counterpart of the canonical partition function. Some formal properties of this superstatistics are 
presented in connection with thermodynamic structures and information aspects. For any combination 
of the local equilibrium statistical factor and the form of fluctuating field, which are ingredients 
of making a generic superstatistics, a variance of the fluctuating quantity appears in the correction term. 
This fact enables us to relate parameters contained in the statistical factor with the variance in 
{\it any} situation. 
\medskip
\end{abstract}
\maketitle
\bigskip
\section{Introduction}
Superposition of statistical factors can be considered as a new attempt to try to formulate a 
statistical mechanical description for unusual systems such as ones having a fluctuating 
temperature field. Since, in the traditional equilibrium statistical mechanics, we suppose 
that systems are put in constant with a heat reservoir (in canonical ensemble), situations 
that arise inhomogeneity in intensive variables over spaces considered may be outside 
the scheme of conventional treatments\cite{Land,Zub}.\\
\indent 
In this paper, we will be particularly concerned with presenting some detailed structure of 
the notion of the superposition of statistical factors and its relation to the fluctuation.
The purpose of this paper is to understand what aspects of a system the superposition of the 
statistical factors delineates. To this end we especially take up of a mixture of the Boltzmann 
factors, which is recently proposed by Beck and Cohen\cite{Beck1}. They named this quantity 
{\it superstatistics}. All considerations through this paper is a situation where local inverse 
temperatures $\beta$, which is usually assumed to be a constant in the ordinary canonical ensemble 
theory, are not homogeneous in space but differ by positions.\\ 
\indent 
The superstatistics denoted by symbol $\mathcal{B}(E)$ allows the infinite types of system's 
distribution in energy $E$, once the fluctuating distribution $f(\beta)$ is given. Three crucial 
premises taken by Beck and Cohen are (a) a system is partitioned into cells that can be considered 
to be reached an equilibrium locally, which is characterized by a single intensive variable (b) 
its statistical factor is Gibbsian (c) the separation between two time scales is adequate, that 
is, the time for relaxing to the each local equilibrium state is much faster than that for varying 
the distribution of $\beta$. Under these assumptions $\mathcal{B}(E)$ is calculated by integrating 
$f(\beta)e^{-\beta E}$ with respect to $\beta$ from zero to infinity. Then the usual Boltzmann 
factor for the system is obtained only when the system has homogeneous temperature 
$\beta=\langle\beta\rangle$ i.e., $f(\beta)=\delta (\beta-\langle\beta\rangle)$. In this view, 
the linear average of the Boltzmann factors over $\beta$ can contain information that characterizes 
the macroscopic behavior.\\ 
\indent
From a formal point of view, Beck and Cohen exemplified that the 
correction to the equilibrium factor $e^{-\langle\beta\rangle E}$ universally follows 
$(1+\sigma^2E^2/2)$, as long as $\sigma E$ is small, where $\sigma$ is the standard deviation 
of $f(\beta)$. Especially when $f(\beta)$ represents the $\chi^2$ distribution the associated 
superstatistics produces the Tsallis type power-law distribution\cite{Wilk,Beck1}, and by 
comparing with the correction term $(1+(q-1)\langle\beta\rangle^2 E^2/2)$ for the $\chi^2$ 
distribution, the parameter $q$ contained in the statistical factor\cite{88Tsallis} can be 
formally related as $q=1+(\sigma/\langle\beta\rangle)^2$. Then the case $q=1$ corresponds 
to no fluctuation in $\beta$, i.e., constant temperature over the system. Once the 
$\mathcal{B}(E)$ is given, the forms of entropy and the associated energy constraints 
can be determined\cite{Tsallis} and the maximum-entropy principle can be a good tool for 
estimating a statistical properties\cite{Sattin}.\\
\indent
As of today, we have taxonomy for the applications of superstatistics (inexhaustive list): 
the log-normal type in turbulence\cite{Beck2,Rey} and the $\chi^2$ 
type in solar flares\cite{Solar}, cosmic rays\cite{Beck4}, and wind velocities in 
airport\cite{Rizzo}. A violent gravitational process can be another possible class of superstatistics, 
where the Gaussian distributions in velocity of stars are equally superposed by 
temperature\cite{Iguc,Chav}. Other recent applications of the superstatistics are briefly 
reviewed in \cite{review}.\\
\indent 
At the present status, however, there seems no guiding principle (dynamical mechanism) to 
choose any specific combination of the forms for the local distribution and the fluctuation 
distributions. We must only refer to the mechanism as a certain complex dynamics and can not 
answer how complex it is. In this sense, the present status on its necessity for introducing 
$\mathcal{B}(E)$ may still be contentious\cite{Lavenda1,Lavenda2}. This ambiguity is also related 
to the fact that some phenomenological data can be fitted for reasonable regions with a 
generalized exponential functions having more than two parameters\cite{comment}, and this fact, 
however, does not imply that the system can be uniquely described by the underlying generalized 
statistical mechanics based on the parameterized entropy which should be maximized.\\
\indent 
These facts are our main reason for avoiding going into discussion of a specific physical system 
in this paper, instead we shall explore some formal connections to the thermodynamic relation 
and to information structure. The plan of this paper is as follows. In Sec.II, we investigate 
six properties possessed by $\mathcal{B}(E)$, which have not been presented so far. The subsections 
A and C are concerned with the connection to thermodynamic structures. From a view point of 
information, we provide a universal relation by considering one dimensional system in subsections D. 
In subsections B, E and F we give results which are direct consequences from the definition but are 
worth presenting in future developments. Sec.III devotes to a discussion for the notion of 
superposition of statistical factors and to a summary. 
\section{Some formal properties of $\mathcal{B}(E)$}
\subsection{Existence of average in $\beta$}\label{secA}
Let us assume that the system considered is situated in an environment where few partitioned region has 
large temperature and the number of regions in the system grows steadily as their temperatures go down. 
The most probable distribution for the regions would be the increasing function of $\beta$. From the 
definition, the superstatistics has the following properties:\\
(i) $\mathcal{B}(E)\rightarrow \infty$ when $E\rightarrow 0$.\\
(ii) $\mathcal{B}(E)$ is a $C^\infty$-class function with respect to $E$ for any $E>0$, i.e., 
\begin{equation}
\mathcal{B}^{(n)}(E)=(-1)^n\int^{\infty}_{min\beta}d\beta\beta^nf(\beta)e^{-\beta E}, \quad n=1,2,\ldots.
\end{equation}
where we set the minimum in $\beta$.\\ 
(iii) $\ln \mathcal{B}(E)$ is a convex function for $E>0$.\\
This fact follows from the direct calculation below:
\begin{eqnarray}
\frac{d^2}{dE^2}\ln \mathcal{B}(E)&=& \frac{\mathcal{B}(E)\mathcal{B}^{''}(E)-(\mathcal{B}^\prime(E))^2}
{\mathcal{B}^2(E)}\nonumber\\
&=& \frac{1}{\mathcal{B}(E)}\int^{\infty}_{min\beta}\left(\beta+\frac{\mathcal{B}^\prime(E)}
{\mathcal{B}(E)}\right)^2e^{-\beta E}f(\beta)d\beta\nonumber\\
&>&0.
\end{eqnarray} 
As an important consequence of the above properties ((i)-(iii)), we can prove the following statement:\\
For $b$ which satisfies $0<min\beta<b$, there exists one single positive solution for any $E>0$ such that 
\begin{equation}
-\frac{d}{dE}\ln \mathcal{B}(E)=b.\label{eqn:Thm}
\end{equation}
\begin{proof}
Putting $\psi (E)=e^{bE}\mathcal{B}(E)$ and using the integrated density of states $F(\beta)$ for the 
temperature distribution, which becomes $F(\beta)\to\infty (\beta\to\infty)$, we write 
\begin{equation}
\psi (E) = \int^{\infty}_{min\beta}e^{-\beta E}dF(\beta).
\end{equation}
Then we can evaluate this as
\begin{eqnarray}
\psi (E) &>& e^{bE}\int^{(min\beta +b)/2}_{min\beta}e^{-\beta E}dF(\beta)
>e^{bE}e^{-E(min\beta +b)/2}\mathcal{I}\nonumber\\
&=&e^{E(b-min\beta)/2}\mathcal{I}
\end{eqnarray}
where $\mathcal{I}=\int^{(min\beta+b)/2}_{min\beta}dF(\beta)$ is a finite. Therefore, we have 
$\psi (E)\to\infty$ when $E\to\infty$. Since the second derivative of $\ln \psi (E)$ 
with respect to $E$ is the same as that of $\ln \mathcal{B}(E)$, the property that 
$\psi (E)\to\infty (E\to 0,E\to\infty)$ together with the property (iii) concludes that 
there should be a single minimum value $E^*$ at which
\begin{equation}
\frac{d \ln\psi (E)}{dE}\bigg|_{E=E^*}=b+\frac{\mathcal{B}^\prime(E^*)}{\mathcal{B}(E^*)}=0 
\end{equation}
is realized.
\end{proof}
Eq.(\ref{eqn:Thm}) can be regarded as a counterpart of the fundamental relation 
$-\partial\ln Z/\partial \beta=E$, which ensures the existence of energy in the system 
in usual statistical mechanics\cite{Khinchin}. We shall find that $b$ represents the average 
inverse temperature in section \ref{disc}.
\subsection{Behavior with low temperature cut-off}
Next, we are concerned with a behavior of the superstatistics at higher temperature. 
If $f(\beta)$ has an asymptotic approximation,
\begin{equation}
f(\beta)\sim a_0\beta^{\gamma_0}+a_1\beta^{\gamma_0}+\cdots +a_n\beta^{\gamma_n}, {\rm as}\quad 
\beta\rightarrow\infty
\end{equation}
where exponents are assumed to to be satisfied with $-1<\gamma_0<\gamma_1<\cdots <\gamma_n$. 
The superposition of the factor with the low temperature cut-off $max\beta$ is defined 
\begin{equation}
\mathcal{B}_{cut}(E)=\int^{max\beta}_{0}f(\beta)e^{-\beta E}d\beta.
\end{equation}
Note that $\mathcal{B}_{cut}(E)\rightarrow \mathcal{B}(E)$ when $max\beta\rightarrow \infty$. 
Then the Watson's lemma\cite{Mathbook} states that 
\begin{equation}
\mathcal{B}_{cut}(E)\sim \sum_{k=0}^na_k\Gamma(\gamma_k+1)E^{-\gamma_k-1},{\rm as} \quad E\rightarrow\infty.
\end{equation}
This states that when $f(\beta)$ consists of some different power laws in $\beta$, the corresponding 
superstatistics is expressed by the superposition of the power laws in energy with different exponents.    
\subsection{Representation by specific heat}
In this section we delineate the superstatistics in the light of a thermodynamic quantity i.e., 
the specific heat of the system. As was supposed in section \ref{secA}, we assume that 
$f(\beta)$ represents the density of states of inverse temperatures in a system around $\beta$, 
which monotonically increases with $\beta$. Then the probability to find a system within the 
interval $\beta$ and $\beta+d\beta$ can be expressed as 
\begin{equation}
p(\beta)d\beta\propto e^{-\beta E}f(\beta)d\beta.
\end{equation}
Since the Boltzmann factor decreases with $\beta$, $p(\beta)$ should have an extremum at a 
certain value of $\beta=\beta^*$, so that 
\begin{equation}
\frac{\partial}{\partial \beta}[e^{-\beta E}f(\beta)]\bigg|_{\beta=\beta^*}=0
\end{equation}
that is 
\begin{equation}
\frac{\partial}{\partial \beta}[\ln f(\beta)]\bigg|_{\beta=\beta^*}=E. 
\end{equation}
Expanding $\ln p(\beta)$ around $\beta^*$, 
\begin{eqnarray}
\ln[p(\beta)]&=&(-\beta^*E+\ln f(\beta^*))\nonumber\\
&+&\frac{1}{2}\frac{\partial^2}{\partial \beta^2}\ln[e^{-\beta E}f(\beta)]
\bigg|_{\beta=\beta^*}(\beta -\beta^*)^2+\cdots\nonumber\\
&\simeq& -\beta^*E+\ln f(\beta^*)-\frac{C_v}{2k_B\beta^*}(\beta-\beta^*)^2.
\end{eqnarray}
where $C_v=(\partial E/\partial T)$ is the specific heat. 
Thus we can evaluate $\mathcal{B}(E)$ as 
\begin{eqnarray}
\mathcal{B}(E)&=&\!\!\int^{\infty}_{0}\!\!P(\beta)d\beta\simeq e^{-\beta^*E+
\ln f(\beta^*)}\!\!\int^{\infty}_{0}e^{-\frac{C_v}{2k_B\beta^*}(\beta-\beta^*)^2}d\beta\nonumber\\
&=& \sqrt{2\pi k_B\beta^*/C_v}e^{-\beta^*E+\ln f(\beta^*)}. 
\end{eqnarray}
Since the specific heat can be usually considered as behaving $O(N)$, $N$ being the number 
of particles in the system, the term which contains $C_v$ contributing as 
$O(\ln N)$ in $\ln\mathcal{B}(E)$ becomes negligible in comparison with others for large $N$. 
In that case $-\beta^*E+\ln f(\beta^*)$ is mainly responsible for the thermodynamic property 
of the system. However we must take into account $C_v$ for systems with small $N$. 
\subsection{$\mathcal{B}(E)$ as an information and its correction by fluctuation}
To illustrate more information aspects on the superstatistics by fluctuation $f(\beta)$, let us 
take a look at another point. The fluctuation distribution can be calculated by an inverse Laplace 
transform with keeping $E$ constant: 
\begin{equation}
f(\beta)=\frac{1}{2\pi i}\int^{c+i\infty}_{c-i\infty}\mathcal{B}(E)e^{\beta E}dE.
\end{equation}
Then the integrand $e^{\ln \mathcal{B}+\beta E}$ tells us that a quantity 
$\phi(\beta,E)=\ln \mathcal{B}(E)+\beta E$ provides information on system's fluctuating field.
Note that this is the same quantity by the Legendre transform of $\mathcal{B}(E)$, which will be 
introduced later in this section. Also, the reader should refer to a discussion revolving around the 
Laplace methods\cite{Mathbook2} for extreme cases $E\to\infty,\beta\to 0$ and 
$E\to 0,\beta\to\infty$\cite{Hugo,Lavenda2}. We instead focus on the evaluation for the intermediate 
energy region. At a saddle point ($E_0$) of the energy coordinate, $\phi(\beta,E)$ should satisfy 
$\phi_0^\prime\equiv (\partial \phi(\beta,E)/\partial E)_{E=E_0}=0$ and 
$\phi_0^{''}\equiv (\partial^2 \phi(\beta,E)/\partial E^2)_{E=E_0}>0$. 
Then when we expand $\phi(\beta,E)$ around $E=E_0$, we have the expression
\begin{equation}
\phi(\beta,E)\sim\phi_0+\frac{1}{2}(E-E_0)^2\phi_0^{''}
\end{equation}
where $\phi_0=\phi(\beta,E_0)$. Therefore we can write 
\begin{equation}
f(\beta)=\frac{e^{\phi_0}}{2\pi i}\int^{c+i\infty}_{c-i\infty}e^{(E-E_0)^2\phi_0^{''}/2}dE.
\end{equation}
A change of variables as $c=E_0$ and $ix=E-E_0 (x\in R)$ provides an expression 
\begin{equation}
f(\beta)=\frac{e^{\phi_0}}{\sqrt{2\pi \phi_0^{''}}}.
\end{equation}
Thus, the corrected {\it bit number}\cite{Beckbook} of the system at equilibrium, 
which is defined as $\ln f(\beta)\equiv I_{MC}(\beta)$, incorporated small 
fluctuations around equilibrium energy $E$, reduces to
\begin{equation}
I_{MC}(\beta)=\phi_0-\frac{1}{2}\ln \phi_0^{''}+\cdots\label{eqn:Imc}.
\end{equation} 
This plays a main role in this study and reminds us the application to the canonical black 
hole entropy by Das {\it et al.}\cite{Das}, where the general corrections to the microcanonical 
entropy of some black holes were calculated. The suffix MC is intended to denote {\it macroscopic} 
because $f(\beta)$ has a global information for the system.
$\phi^{''}(\beta,E_0)$ in the above equation is found to be equivalent to 
$\langle \beta^2\rangle_s-\langle \beta\rangle_s^2$ if we employ the average of $\beta$ and 
$\beta^2$ by means of $p_s$ as we will see in the next section, which allows us to express as 
$I_{MC}(\beta)\simeq \phi_0-\ln [\langle \beta^2\rangle_s-\langle \beta\rangle_s^2]/2$.\\
\indent 
In order to delineate further structure which is universally possessed by $\mathcal{B}(E)$, we consider 
one dimensional system whose segment at position $n$ is assumed to be in a local equilibrium prescribed by 
a Boltzmann factor $e^{-\beta_n E}$. In order to see the effect of the fluctuation of temperature on the bit 
number of whole system, we can borrow an approach that was taken by \cite{BH} for the canonical entropy of 
very large black holes obeying the Beckenstein-Hawking area law\cite{area}. In this approach, 
it was shown to have the universal logarithmic correction to the microcanonical black hole entropy. 
The superstatistics for the present system may be expressed as 
\begin{equation}
\mathcal{B}(E)=\sum_{n=-\infty}^\infty f(\beta_n)e^{-\beta_n E}.
\end{equation}
We would be appropriate to interpret this form as weighted sum of the local factors according to the value of 
inverse temperature of the segment $n$. 
We can move to a continuous expression 
\begin{equation}
\mathcal{B}(E)\simeq \int^{\infty}_{-\infty}f(\beta (x))e^{-\beta (x) E}dx,
\end{equation}
where we have applied the Poisson resummation formula, 
\begin{equation}
\sum_{n=-\infty}^\infty g(n)=\sum_{m=-\infty}^\infty\int^{\infty}_{-\infty}dx g(x)e^{-i2\pi mx} 
\end{equation}
which holds for sufficiently large $x$ with $\sum_m e^{-i2\pi mx}\approx 1$.  
Then by converting to the integration with respect to $\beta$ we obtain an expression 
\begin{eqnarray}
\mathcal{B}(E)&=& \int^{\infty}_{0}d\beta \left( \frac{d\beta}{dx}\right)^{-1}f(\beta)e^{-\beta E}\nonumber\\
&\simeq& \int^{\infty}_{0}\exp[I_{MC}(\beta)-\beta E -\ln\left( \frac{d\beta}{dx}\right)]d\beta.
\end{eqnarray}
By employing the method of steepest descent for the saddle point $\beta_0$, we have 
\begin{eqnarray}
\mathcal{B}(E)&\simeq& \left(\frac{2\pi}{-I_{MC}^{''}(\beta_0)}\right)^{\frac{1}{2}}\nonumber\\
&\times& \exp\left[I_{MC}(\beta_0)-\beta_0 E -\ln\left( \frac{d\beta}{dx}\right)\bigg|_{\beta=\beta_0}\right].
\end{eqnarray}
Here we define $I_C(\beta)$ by the Legendre transform of pairs 
$(E,-\ln \mathcal{B}(E)) \rightarrow (\beta,I_C(\beta))$, i.e., 
\begin{equation}
I_C(\beta)=\ln \mathcal{B}(E)+\beta E,
\end{equation}
which is the same as $\phi(\beta,E)$ introduced in the beginning of this section. 
This is an analogy of the standard thermodynamic definition of canonical entropy $S_c$ 
from the canonical partition function $Z_c$; $-\ln Z_c=-S_c +\beta E$. Then $I_C(\beta)$ 
can be interpreted as a quantity which contains information of fluctuation and spacial 
inhomogeneity. Therefore $I_C(\beta)$ evaluated at the saddle point $\beta=\beta_0$ is 
calculated to be 
\begin{eqnarray}
I_C(\beta_0)&=&I_{MC}(\beta_0)-\ln\left(\frac{d\beta}{dx}\right)\bigg|_{\beta=\beta_0}-\frac{1}{2}
\ln[-I_{MC}^{''}(\beta_0)]\nonumber\\
&+&\frac{1}{2}\ln(\frac{\pi}{2}).
\end{eqnarray}
The correction term $I_C(\beta_0)-I_{MC}(\beta_0)\equiv \Delta I$ is provided by 
\begin{equation}
\Delta I=-\frac{1}{2}\ln\left[-\frac{2}{\pi}\frac{d^2 I_{MC}}{d\beta^2}\left(\frac{d\beta}{dx}\right)^2
\bigg|_{\beta=\beta_0}\right].\label{eqn:shift}
\end{equation}
This formula is expected to hold for the most general circumstances without depending on particular 
choice of $f(\beta)$ and $\beta(x)$. However observe, for example, that with a constant temperature 
gradient in $x$, which is a possible setting for systems, the quantity in the square brackets 
in Eq.(\ref{eqn:shift}) can be negative for some $f(\beta)$\cite{exmp}. Moreover, a uniform 
distribution and finite level distributions always give $d^2I_{MC}/d\beta^2=0$. 
This explains that the saddle-point approximation for the calculation of the effect of 
fluctuation (the logarithmic shift term) to the macroscopic bit number is limited to the case 
where $I_{MC}^{''}(\beta_0)\le 0$ is satisfied.  
\subsection{Discrete levels for $\mathcal{B}^{-1}(E)$ without quantization}\label{disc} 
In this section, we seek yet another connection to thermodynamic expression and the superstatistics.  
What we are concerned is a relation with the variance of $\beta$. To see this, we introduce a 
conjugate distribution function\cite{Khinchin}
\begin{equation}
p_s=\frac{f(\beta)e^{-\beta E}}{\mathcal{B}(E)};\int^{\infty}_{0}p_sd\beta=1.\label{eqn:cnj}
\end{equation}
Then 
\begin{eqnarray}
\frac{d^2\ln \mathcal{B}(E)}{dE^2}&=&\frac{\mathcal{B}^{''}(E)}{\mathcal{B}(E)}-
\left(\frac{\mathcal{B}^\prime(E)}{\mathcal{B}(E)}\right)^2 \nonumber\\
&=&\langle \beta^2\rangle_s-\langle \beta\rangle_s^2
\end{eqnarray}
where $\langle \cdot\rangle_s=\int^{\infty}_{0}(\cdot)p_sd\beta$.
On the other hand, since $\langle \beta\rangle_s=-(\partial \ln \mathcal{B}(E)/\partial E)$, 
\begin{eqnarray}
\frac{\partial \langle \beta\rangle_s}{\partial E}&=& \frac{\partial}{\partial E}\left[ 
\frac{1}{\mathcal{B}(E)}\int^{\infty}_{0}\beta f(\beta)e^{-\beta E}d\beta\right]\nonumber\\
&=& -\langle \beta^2\rangle_s + \langle \beta\rangle_s^2\label{eqn:disc}
\end{eqnarray}
therefore we obtain a thermodynamical relation 
\begin{equation}
\frac{\partial^2 \ln \mathcal{B}(E)}{\partial E^2}=-\frac{\partial \langle \beta\rangle_s}{\partial E}.
\end{equation}
From Eq.(\ref{eqn:disc}), we obtain a discrete expression for $\mathcal{B}^{-1}(E)$ as follows.

Putting $\langle \beta^2\rangle_s=\epsilon^2$, 
\begin{equation}
dE=\frac{d\langle \beta\rangle_s}{\langle \beta\rangle_s^2-\epsilon^2}.
\end{equation}
It follows that  
\begin{equation}
2\epsilon dE=\left(\frac{1}{\langle \beta\rangle_s-\epsilon}-\frac{1}{\langle \beta\rangle_s+\epsilon}
\right)d\langle \beta\rangle_s.
\end{equation}
Solving this leads to 
\begin{equation}
\frac{\langle \beta\rangle_s-\epsilon}{\langle \beta\rangle_s+\epsilon}e^{-2\epsilon E}=c
\end{equation}
where $c$ is an integral constant. The assumption that the system's temperature on average 
would approach to zero if its energy vanishes by a classical description (i.e., 
$\langle \beta\rangle_s \to \infty$ when $E\to 0$) provides $c=1$. Then the solution becomes
\begin{equation}
\langle \beta\rangle_s=-\epsilon\coth(\epsilon E).
\end{equation}
We have 
\begin{equation}
\ln \mathcal{B}(E)=\int \epsilon\coth(\epsilon E)=\ln[a\sinh(\epsilon E)]
\end{equation}
where $a$ is a constant of integration. Therefore
\begin{eqnarray}
\mathcal{B}^{-1}(E)&=&\frac{2}{a(e^{\epsilon E}-e^{-\epsilon E})}\nonumber\\
&=&\frac{2e^{-\epsilon E}}{a}(1+e^{-2\epsilon E}+e^{-4\epsilon E}+\cdots)\nonumber\\
&=&\frac{2}{a}\sum_{n=0}^\infty e^{\mathcal{E}_n}
\end{eqnarray}
where $\mathcal{E}_n=2 (n+\frac{1}{2})\sqrt{\langle \beta^2\rangle_s}E$ is discrete levels 
for the inverse of the superstatistics with the lowest $\sqrt{\langle \beta^2\rangle_s}E$. 
This discrete quantity has a dimension of specific heat, so that we can view that the 
system can be characterized by a superposition of many different levels of hierarchies with 
different classes of specific heat. However note that they have a common energy $E$. 
\subsection{An inequality for $\mathcal{B}(E)$}
We consider a situation where single $\beta$ in a Gibbsian statistical factor is replaced by more 
generic one. The sum of these factors which can be regarded as a discrete expression for the 
superstatistics may be written as 
\begin{equation}
\mathcal{B}(E)=\sum_{\{\beta\}}e^{-E\zeta(\{\beta\})}
\end{equation}
where $\zeta(\{\beta\})$ is a quantity which is uniquely determined in a system once a 
distribution of $\beta^{-1}$'s is given. Let us consider a bound which is satisfied by 
$\mathcal{B}(E)$ from a perturbation theory about the distribution of the inverse temperatures 
over the system. We express the difference in $\zeta(\{\beta\})$ between a certain realization 
of $\beta^{-1}$ distribution over a system and its coarse-grained one as $\Delta\zeta(\{\beta\})$ 
(i.e., $\Delta\zeta(\{\beta\})=\zeta(\{\beta\})-\zeta_c(\{\beta\})$). By the coarse-grained inverse 
temperature field we mean that the system has a single homogeneous one over the system and it 
appears to be reached at an equilibrium states if we look at whole system from far away. 
If we look closer at the system we should see the fluctuating field. This illustration also 
relates to the number of division of the whole system considered. The role of $\beta_i$'s in 
the exponential function is just the same as the spin variables $\{s_i\}$ in the Ising Hamiltonian 
system, where the partition function and the system's energy are quantities for discussing the 
theory instead of $\mathcal{B}(E)$ and $\zeta$ respectively. Therefore we can write 
\begin{equation}
\mathcal{B}(E)=\sum_{\{\beta\}}e^{-E(\zeta_c(\{\beta\})+\Delta\zeta)}
=\mathcal{B}_c(E)\langle e^{-E\Delta\zeta}\rangle_c
\end{equation}
where we have put the average as
\begin{equation}
\langle\cdot\rangle_c=\sum_{\{\beta\}}(\cdot)e^{-E\zeta_c}/\mathcal{B}_c(E),\quad \mathcal{B}_c(E)=
\sum_{\{\beta\}}e^{-E\zeta_c}.
\end{equation}
Under the condition that the deviation is small i.e., $\Delta\zeta\ll 1$,
\begin{eqnarray}
\langle e^{-E\Delta\zeta}\rangle_c & = & 1-E\langle \Delta\zeta\rangle_c+\cdots\nonumber\\
& \simeq & e^{-E\langle \Delta\zeta\rangle_c}.
\end{eqnarray}
Thus, within the first order perturbation, $\mathcal{B}(E)$ can be expressed as
\begin{equation}
\mathcal{B}(E)\simeq \mathcal{B}_c(E) e^{-E\langle \zeta-\zeta_c\rangle_c}.
\end{equation}
Furthermore, since an inequality $\langle e^f\rangle\ge e^{\langle f\rangle}$\cite{ineq}, 
we can obtain an inequality which corresponds to the Gibbs-Bogoliubov-Feynmann inequality\cite{GBF} 
in the usual statistical mechanics
\begin{equation}
\mathcal{B}(E)\ge \mathcal{B}_c(E) e^{-E\langle \zeta-\zeta_c\rangle_c}.
\end{equation}
It is worth noting that when $\zeta$ is simply equivalent to $\beta$ the equality holds and 
$\mathcal{B}(E)$ provides the Boltzmann factor, which corresponds to the environment with no 
fluctuation in $\beta$.   
\section{Discussion and Summary}
In this paper, we have investigated some thermodynamical structures and information aspects of 
the superposition of statistical factors which is achieved by the assumption of the Gibbsian 
local equilibrium distribution and the existence of the spacial fluctuation of the inverse 
temperature over the system. We spotlighted some structural side of the superstatistics rather 
than the concrete application of it to the physical system.\\
\indent 
To emphasize the noticeable similarity in the structure to the canonical partition function, 
we can take a standpoint of the generating function\cite{Khinchin}, which is constructed by 
a product of a statistical factor and a structure function $\omega(x)$ (density of states). 
The generating function $G(\mu)=\int^{\infty}_{0}\omega(\mu)e^{-\mu\nu}d\mu$ can be related 
to the variance of its variables as 
\begin{equation}
G(\mu)\simeq e^{-\mu\langle\nu\rangle}\left(1+\frac{1}{2}(\sigma_\nu\mu)^2\right)\label{eqn:Gmu}
\end{equation}
where $\sigma_\nu^2=\langle\nu^2\rangle-\langle\nu\rangle^2$ is a variance calculated from 
a definition $\langle \cdot\rangle=\int^{\infty}_{0}(\cdot)\omega(\nu)d\nu$. $\mu$ and $\nu$ 
represent either $E$ or $\beta$ respectively. When the symbols denote $(\mu,\nu)=(E,\beta)$ 
and $(\beta,E)$, the generating function represents $Z_c(\beta)$ and $\mathcal{B}(E)$ 
respectively. From this observation, we arrived at a view that the canonical partition 
function and superstatistics have a universal relation to the degree of dispersion of 
environmental variables irrespective of the form of density of states (for $E$ and $\beta$), 
indicating that they are two different aspects of the single generating function (Table.I).
\begin{table}[h]
\caption{\label{I} Gibbsian superstatistics and partition function}
\begin{tabular}{@{}l|c|c}
\hline
Statistical factor & Density of states & Generating function\\
\hline
Gibbsian & $\rho (E)$ & $Z_c(\beta)$ \\
$e^{-\beta E}$ & $f(\beta)$ & $\mathcal{B}(E)$ \\
\hline
\end{tabular}
\end{table}
\indent 
In order to extract underlying structure of notion of superposition of statistical factors, we 
here study more generic circumstance for a statistical factor $g(\beta,E)$, which is assumed 
to be locally realized. We write $g(\beta,E)=g(\langle\beta\rangle,E)\xi(\langle\beta\rangle,\beta,E)$, 
where $\xi(\langle\beta\rangle,\beta,E)=g^{-1}(\langle\beta\rangle,E)g(\beta,E)$. Therefore 
the superstatistics of the general statistical factor $g$ becomes 
\begin{equation}
\mathcal{B}(E)=\int^{\infty}_{0}\omega(\beta)g(\beta,E)d\beta=g(\langle\beta\rangle,E)
\langle\xi(\langle\beta\rangle,\beta,E)\rangle.
\end{equation}
By Taylor expanding $\xi(\langle\beta\rangle,\beta,E)$ around $\langle\beta\rangle$, we obtain 
\begin{equation}
\mathcal{B}(E)=g(\langle\beta\rangle,E)\left[1+\frac{1}{2}\sigma^2
\frac{d^2\xi}{d\beta^2}\bigg|_{\beta=\langle\beta\rangle}+\cdots\right].
\end{equation}
This allows us to conclude that it is possible to relate parameters contained in the distribution 
$\omega(\beta)$ to the temperature dispersion $\sigma^2$ for {\it any} statistical factors. 
Beck and Cohen showed that when $\omega(\beta)$ is the gamma distribution and $g=e^{-\beta E}$ 
an index $q$ appearing in the Tsallis factor can be expressed as 
$q=\langle\beta^2\rangle/\langle\beta\rangle^2$\cite{Beck1}. However, as we demonstrated above, 
this feature applies universally for the other combination of superposition of statistical 
factors\cite{comb} once we assume the local equilibrium states for each partitioned region. 
We give two examples when the local equilibrium does not obey the Gibbsian form. 
In the cases of a Tsallis type power-low and a stretched exponential type local equilibrium, 
we respectively have 
\begin{eqnarray}
\mathcal{B}(E) \simeq e_q^{-\langle\beta\rangle E}\left(1+\frac{1}{2}q\sigma^2E^2\right)\nonumber
\end{eqnarray}
where $e_q^x$ is the $q$-exponential function frequently used in the nonextensive statistical 
mechanics\cite{88Tsallis,NSM} when $\sigma E$ is not large and  
\begin{eqnarray}
\mathcal{B}(E) \simeq e^{-(\langle\beta\rangle E)^\eta}\left( 1+\frac{1}{2} y(\eta)\sigma^2\right)\nonumber
\end{eqnarray}
where $y(\eta)=-\eta E\eta\langle\beta\rangle^{\eta-2}(\eta-\eta E^\eta\langle\beta\rangle-1)$.
Then the universal quadratic correction terms to the alternative factors   
$1+q\sigma^2E^2/2$ and $1+y(\eta)\sigma^2/2$ 
are respectively needed no matter what the precise form of the fluctuation is. 
Note that we have the same form as Eq.(\ref{eqn:Gmu}) when $q=1$ and $\eta=1$ 
because the statistical factors recover the Gibbsian.\\
\indent 
Although the superstatistics constructed in the sense of \cite{Beck1} plays a role of a new 
distribution incorporated by system's global information on fluctuating inverse temperature, 
it is not normalizable with respect to energy $E$. The situation is true for a partition function 
$Z(\beta)$ in the usual statistical mechanics, which gives $\int^{\infty}_{0}Z(\beta)d\beta\neq 1$. 
We may construct a conjugate distribution function Eq.(\ref{eqn:cnj}), allowing us to write the 
field $f(\beta)=\mathcal{B}(E)e^{\beta E}p_s$ for the purpose of explaining (fitting) observed 
distribution functions. A quantity $\ln \mathcal{B}^{-1}(E)/E$ can capture states of the system and can be 
finite even when $E\to\infty$\cite{IC} due to concavity of $\ln \mathcal{B}^{-1}(E)$. In this paper, 
we did not consider the dynamical mechanism to keep the fluctuation $f(\beta)$ as a specific form. 
Instead, we would like to mention what we can say more about a relation between variance $\sigma$ 
and a thermodynamic force (affinity) when we regard $F(t)\equiv -\ln\mathcal{B}(E)/E$ 
as a fundamental quantity to see the approaching process to uniform temperature over the system: 
\begin{eqnarray}
\Delta F(t)=F(t)-F_0=-\frac{1}{E}\ln\left(\frac{\mathcal{B}_t(E)}{\mathcal{B}_0(E)}\right)\nonumber
\end{eqnarray}
where $\mathcal{B}_t(E)$ is a superstatistics calculated from $f(\beta)$ at time $t$ and 
$\mathcal{B}_0(E)$ denotes one that from uniform temperature. $\mathcal{B}_0(E)$ is equivalent 
to the factor $e^{-\langle\beta\rangle E}$. Then, in consideration of the universal correction 
term derived for all types of $f(\beta)$, up to the second order in energy, $\Delta F(t)$ is 
found to be $\Delta F(t)=-\ln\left(1+\frac{1}{2}\sigma^2(t)E^2\right)/E$, or equivalently 
$\sigma(t)=\sqrt{2(e^{-E\Delta F(t)}-1)}/E$, expressing that 
the system has a single temperature in the course of the thermodynamics force vanishes.\\
\indent 
In summary, we have explored some formal properties of the superposition of statistical factors 
proposed by Beck and Cohen. To understand what the notion of superstatistics represents, 
we have attempted to see links to information and thermodynamic relation. Non-equilibrium states 
or relaxation processes to uniform temperature field over the system would be interesting 
for the future research.  
\acknowledgements
The author acknowledges M. Morikawa for discussion and the JSPS fellowship No.06225 with 
support in part from the Grant-in-Aid for Scientific Research from the Ministry of Education, 
Science, Sports and Culture of Japan.


\begin{thebibliography}{99}
\bibitem{exmp} An example is the log-normal distribution 
$f(\beta)=(\sqrt{2\pi}\beta s)^{-1}e^{-\log(\beta/m)^2/2s^2}$, where $s$ and $m$ are parameters, 
which gives $d^2I_{MC}/d\beta^2=\beta^{-2}(1+s^{-2})>0$. On the contrary, the Gaussian distribution 
gives $d^2I_{MC}/d\beta^2=-\sigma^{-1}$, where $\sigma^2=\langle \beta^2\rangle -\langle \beta\rangle^2$, 
then Eq.(\ref{eqn:shift}) is usable. The same discussion is true for the other combinations of 
$f(\beta)$ and $\beta(x)$.
\bibitem{ineq} This is a direct consequence of an inequality $e^x\ge 1+x$.
\bibitem{GBF} The inequality is expressed as $Z(\beta)\ge Z_{MF}(\beta)e^{-\beta\langle E-E_{MF}\rangle_{MF}}$ 
where the suffix MF denotes the mean field theory\cite{Chandler}.
\bibitem{comb} For example, the superstatistics for the system with the Tsallis local equilibrium 
states when the fluctuating inverse temperature field obeys a power low with index $z$ corresponds 
to the Mellin transform of the Tsallis factor $e_q^x$, i.e., $\mathcal{M}\{e_q^x;E\}$. Its exact 
expression can be calculated both for $q>1$ and $0<q<1$\cite{TY}.
\bibitem{IC} Note however that if $\mathcal{B}(E)$ is normalizable we would provide a picture 
of information content per unit energy, but this does not apply in this case.
\end{thebibliography}
\end{document}